\documentclass[12pt]{iopart}

\usepackage{graphicx}
\usepackage{subfig}
\usepackage{multirow}
\usepackage{booktabs}
\usepackage{xcolor}
\usepackage{iopams}  

\begin{document}

\title[]{Application of Quantum Machine Learning in a Higgs Physics Study at the CEPC}

\author{Abdualazem Fadol $^{1,5}$, Qiyu Sha $^{1,2}$, Yaquan Fang $^{1,2}$, Zhan Li $^{1,2}$, Sitian Qian $^{3}$, Yuyang Xiao $^{3}$, Yu Zhang $^{4}$, Chen Zhou $^{3,}$*}

\address{%
$^{1}$ Institute of High Energy Physics, 19B Yuquan Road, Shijingshan District, Beijing 100049, China\\
$^{2}$ University of Chinese Academy of Sciences, 19A Yuquan Road, Shijingshan District, Beijing 100049, China\\
$^{3}$ State Key Laboratory of Nuclear Physics and Technology, School of Physics, Peking University, 209 Chengfu Road, Haidian District, Beijing 100871, China\\
$^{4}$ Qujing Normal University, 222 Sanjiang Road, Qilin District, Qujing 655011, Yunnan Province, China\\
$^{5}$ Spallation Neutron Source Science centre, Dongguan 523803, China
}
\ead{$^*$czhouphy@pku.edu.cn}
\vspace{10pt}

\begin{abstract}
Machine learning has blossomed in recent decades and has become essential in many fields. It significantly solved some problems in particle physics---particle reconstruction, event classification, etc. However, it is now time to break the limitation of conventional machine learning with quantum computing. A support-vector machine algorithm with a quantum kernel estimator (QSVM-Kernel) leverages high-dimensional quantum state space to identify a signal from backgrounds. 
In this study, we have pioneered employing this quantum machine learning algorithm to study the $e^{+}e^{-} \rightarrow ZH$ process at the Circular Electron-Positron Collider (CEPC), a proposed Higgs factory to study electroweak symmetry breaking of particle physics. Using 6 qubits on quantum computer simulators, we optimised the QSVM-Kernel algorithm and obtained a classification performance similar to the classical support-vector machine algorithm. Furthermore, we have validated the QSVM-Kernel algorithm using 6-qubits on quantum computer hardware from both IBM and Origin Quantum: the classification performances of both are approaching noiseless quantum computer simulators. In addition, the Origin Quantum hardware results are similar to the IBM Quantum hardware within the uncertainties in our study. 
Our study shows that state-of-the-art quantum computing technologies could be utilised by particle physics, a branch of fundamental science that relies on big experimental data.
\end{abstract}

\vspace{2pc}
\noindent{\it Keywords}: Quantum computing, Machine Learning, Particle Physics, Higgs Factory
%
%
%
%

\section{Introduction}

The discovery of the Higgs boson~\cite{atlashiggs,cmshiggs} by the ATLAS and CMS experiments at the Large Hadron Collider (LHC) in 2012 was a significant milestone in particle physics. It confirmed the fundamental particle spectrum of the Standard Model and opened a new window to refine our understanding of particle physics. The Higgs boson is needed to break the electroweak symmetry in the Standard Model. Since then, the LHC experiments have performed extensive studies on the Higgs boson properties: clues for new physics would emerge if any measurement disagrees with the Standard Model prediction. However, there is no significant hint of new physics has been found to date. Higgs factories~\cite{ILC,FCC,CLIC,CEPC-SPPC} based on lepton colliders have been proposed to perform more precise measurements of the Higgs boson properties and study electroweak symmetry breaking of particle physics. The Circular Electron-Positron Collider (CEPC), presented by Chinese scientists, is one such collider that acts as a Higgs factory. It will be located in a tunnel with a circumference of approximately 100 km colliding electron-positron pairs at a centre-of-mass energy of up to $240\ \mathrm{GeV}$, upgradable to $360\ \mathrm{GeV}$. 

Machine learning has enjoyed widespread success in detector simulation, particle reconstruction and data analyses of experimental particle physics and dramatically enhances the ability to achieve physics discovery. For instance, machine learning algorithms are used in ATLAS and CMS experiments to help separate signals from backgrounds in the observation of the Higgs boson production in association with a top quark pair ($t\bar{t}H$), which directly establishes the Higgs boson couplings to the top quarks~\cite{atlastth,cmstth}. 

Another essential tool for experimental particle physics could be quantum machine learning. It uses quantum computing to perform machine learning tasks that tackle large data dimensions. Quantum machine learning enables effective operations in high-dimensional quantum state spaces where computers operate with qubits instead of classical bits. Therefore, it could provide fast computing speed and better learning ability than classical machine learning. As an example of quantum machine learning, a support-vector machine algorithm with a quantum kernel estimator (QSVM-Kernel)~\cite{qsvmv,qsvmm} encodes classical data into quantum state space and makes accurate classifications for certain artificial data sets.

In recent years, the field of quantum computing has developed rapidly. Superconducting quantum and optical quantum computers have been successfully fabricated and have demonstrated capabilities far beyond today's supercomputers in certain computing tasks~\cite{qagoogle,qaustc}. In the following decades, this field will likely increase the number of qubits, improve execution time, and reduce device noise for quantum computers. These developments will lay a foundation for the practical application of quantum computing.

Studying quantum machines to utilise the potential of quantum computing advantage for future particle physics research is important. There have already been proof-of-principle studies that apply quantum machine learning algorithms to detector simulation, particle reconstruction and data analyses. For example, the QSVM-Kernel algorithm has been employed in a few physics analyses at the LHC (such as the $t\bar{t}H$ measurement) using quantum computer simulators and superconducting quantum computer hardware~\cite{qsvmtth,qml4hep1,qml4hep2}. These studies confirm that quantum machine learning algorithms have the ability to separate signals from backgrounds for certain realistic physics data sets. However, further improvements in both quantum algorithms and quantum devices are still required before using quantum machine learning in particle physics experiments. 

In this study, we focus on applying the QSVM-Kernel algorithm to a physics analysis that measures $ZH$ (the Higgs boson production in association with a $Z$ boson) at the CEPC. Using quantum computer simulators, we pursue quantum algorithm designs that are more suitable for particle physics data analyses. On the other hand, we validate the performance of the QSVM-Kernel algorithm using superconducting quantum computer hardware provided by IBM and Origin Quantum, a quantum technology company in China. 

\section{Materials and Methods}
\subsection{Physics Analysis}
The CEPC will operate at a centre-of-mass energy of about $240\ \mathrm{GeV}$ where the Higgs boson production cross-section reaches maximum through the $e^{+} e^{-} \rightarrow Z H$ process. Over one million Higgs bosons will be produced with an integrated luminosity of $5.6\ \mathrm{ab}^{-1}$. These large statistics and the clean final states of electron-positron collisions will allow the CEPC experiments to measure the Higgs boson properties with precision far beyond that at the LHC~\cite{CEPC_whitepaper}.
$e^{+}e^{-} \rightarrow ZH$ events can be tagged using the mass of the system recoiling against the $Z$ boson (``recoil mass''), calculated using the difference of the four-momentum between the electron-positron pair and $Z$ boson. Combining this method with measurements in individual Higgs decay channels will allow for model-independent measures of both $e^{+}e^{-} \rightarrow ZH$ production cross section and Higgs decay branching ratios, which is not feasible at the LHC. These measurements will extract information on Higgs boson couplings to other fundamental particles and provide sensitive probes to new physics beyond the Standard Model.

\begin{figure}[!t]
  \begin{center}
  \includegraphics[width=0.5\textwidth]{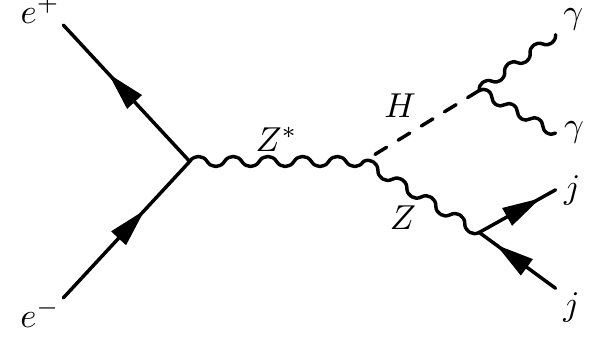}
  \end{center}
  \caption{Representative Feynman diagrams for the $e^{+}e^{-} \rightarrow ZH \rightarrow jj\gamma \gamma$ production.}\label{fig_feynman_diagrams}
\end{figure}

The study focuses on the $H \rightarrow \gamma \gamma, Z \rightarrow jj$ decay mode of the $e^{+}e^{-} \rightarrow ZH$ process, where $\gamma$ and $j$ represent a photon and a jet, respectively. Figure~\ref{fig_feynman_diagrams} shows a representative Feynman diagram for this process. Although the diphoton decay of the Higgs boson has a small branching ratio in the SM, the two photons can be well identified and measured, which boosts the precision in this channel. Events are required to have two photon candidates retained as the $H \rightarrow \gamma \gamma$ candidate and two jets retained as the $Z \rightarrow jj$ candidate. The leading photon candidate is required to have energy greater than $35 \ \mathrm{GeV}$, the sub-leading photon candidate is required to have energy greater than $25 \ \mathrm{GeV}$, and both are required to have polar angles of $|\cos\theta| < 0.9$. In this analysis, jets are selected based on their polar angles, with a requirement of $|\cos\theta| < 0.9$. Furthermore, the main SM background process is the $e^{+}e^{-} \rightarrow (Z/\gamma^{*}) \gamma\gamma$ process, in which the photons ($\gamma$'s) originate from either initial or final state radiation.

The Higgs signal and the SM background events are generated with \textsc{WHIZARD-1.95} \cite{WHIZARD} and fast simulated with the CEPC baseline detector design~\cite{CEPC-SPPC}. One can find more details about the event generation on Ref~\cite{event_generation}.

\begin{figure}[ht]
  \begin{center}
  \subfloat[\label{subfig:DeltaP_yy_jj}]{\includegraphics[width=0.33\textwidth]{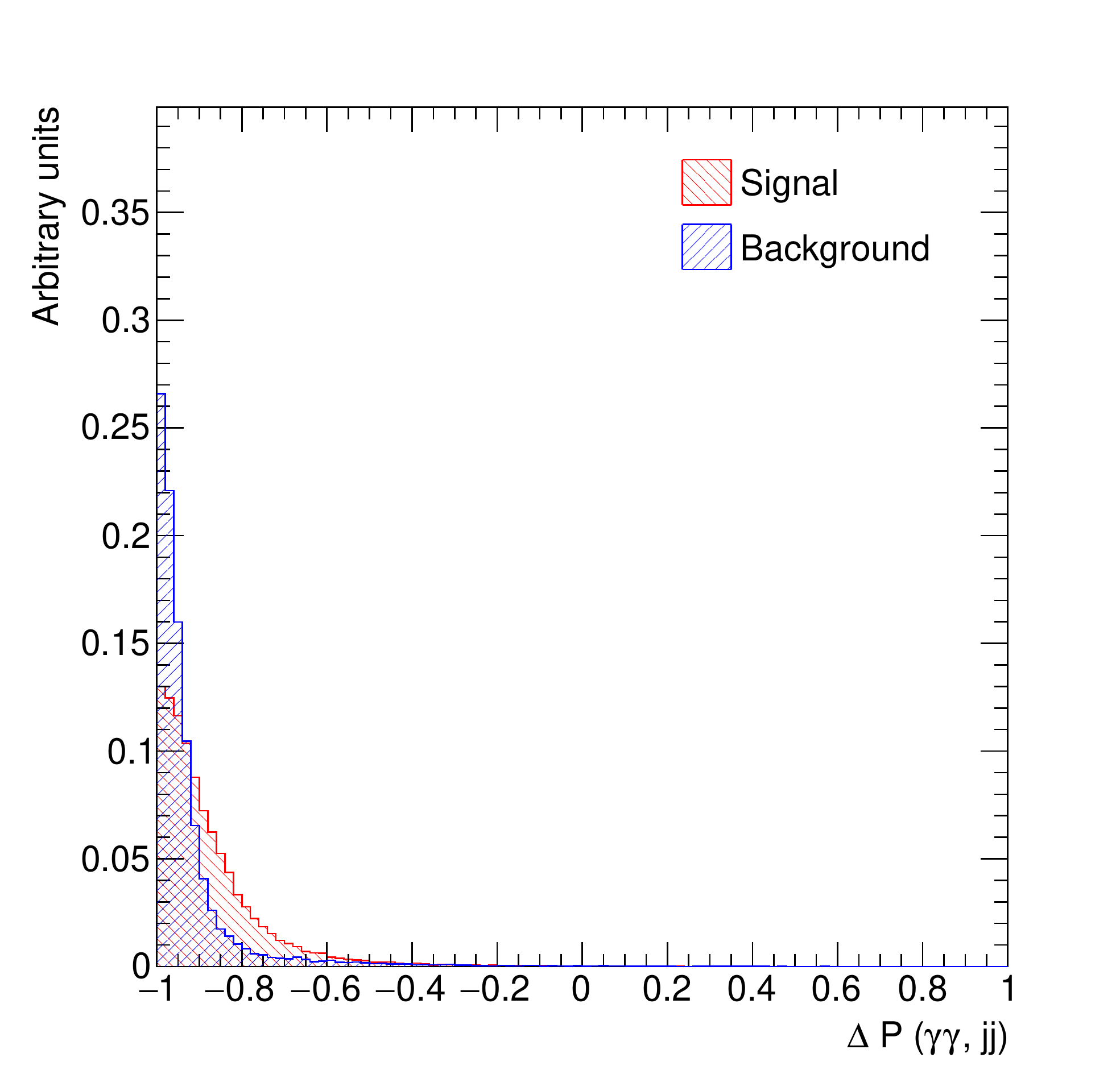}}
  \subfloat[\label{subfig:DeltaPhi_yyy}]{\includegraphics[width=0.33\textwidth]{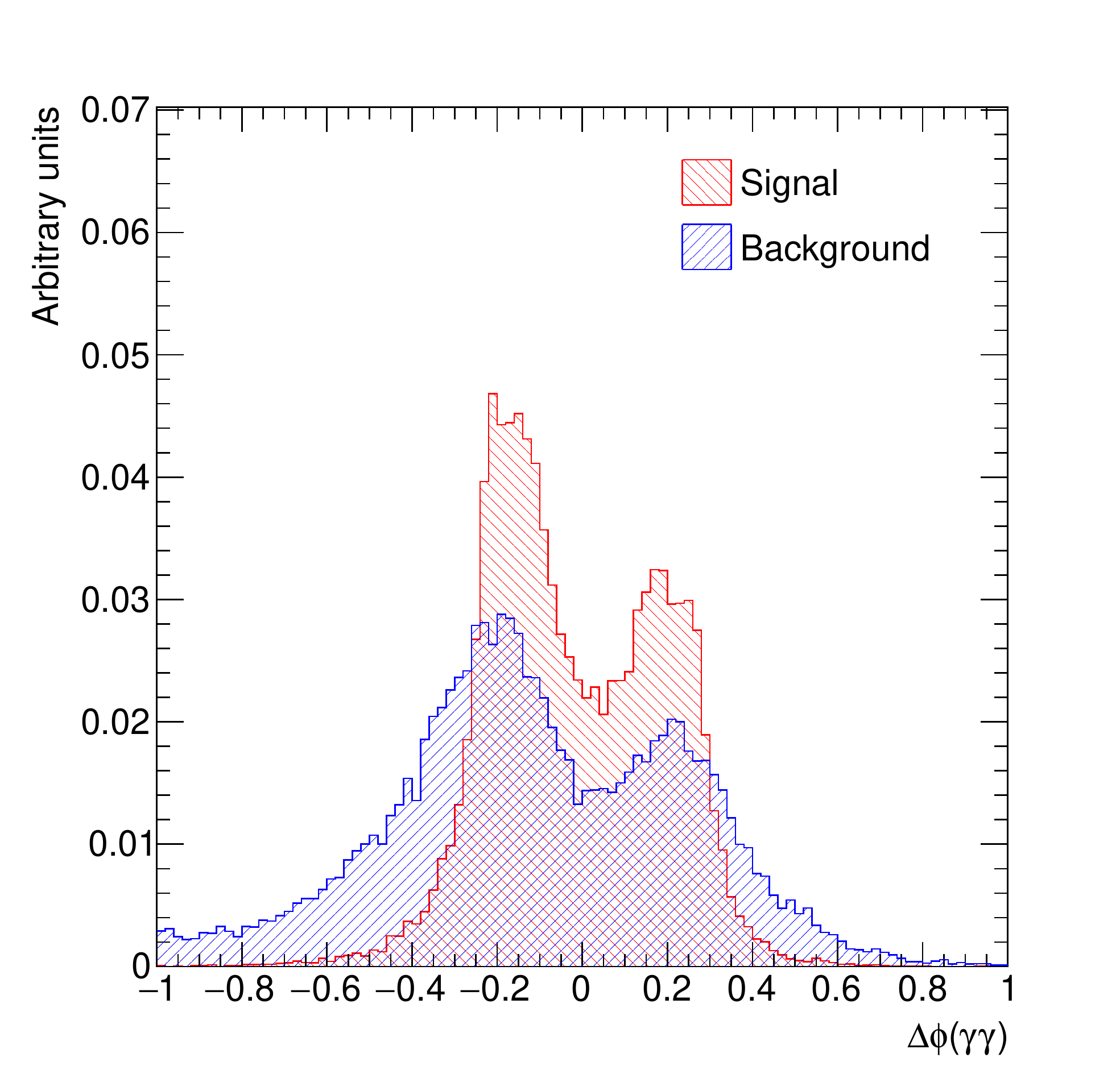}}
  \subfloat[\label{subfig:minDeltaR_y_j}]{\includegraphics[width=0.33\textwidth]{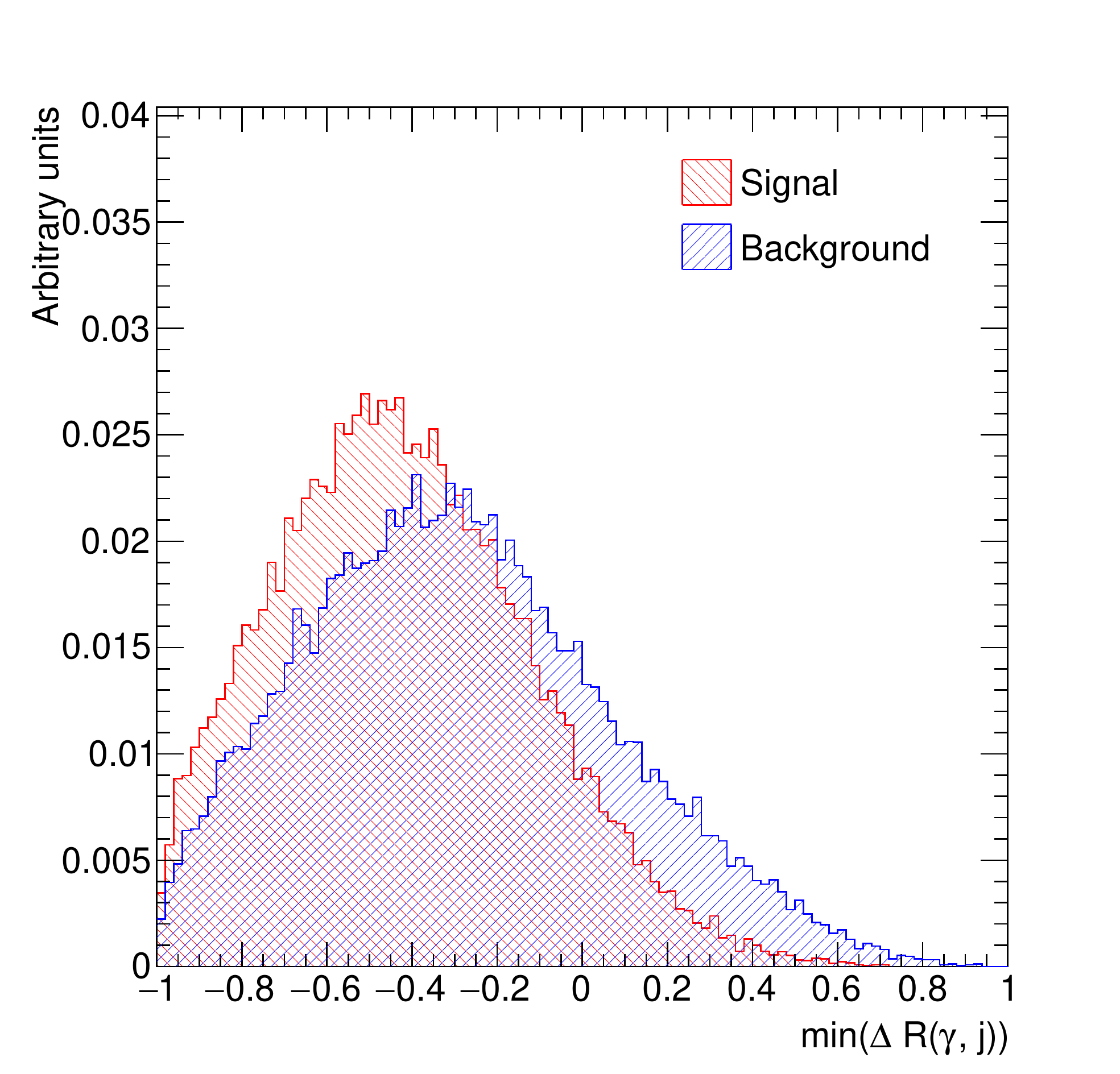}}\\
  \subfloat[\label{subfig:recoM_jj}]{\includegraphics[width=0.33\textwidth]{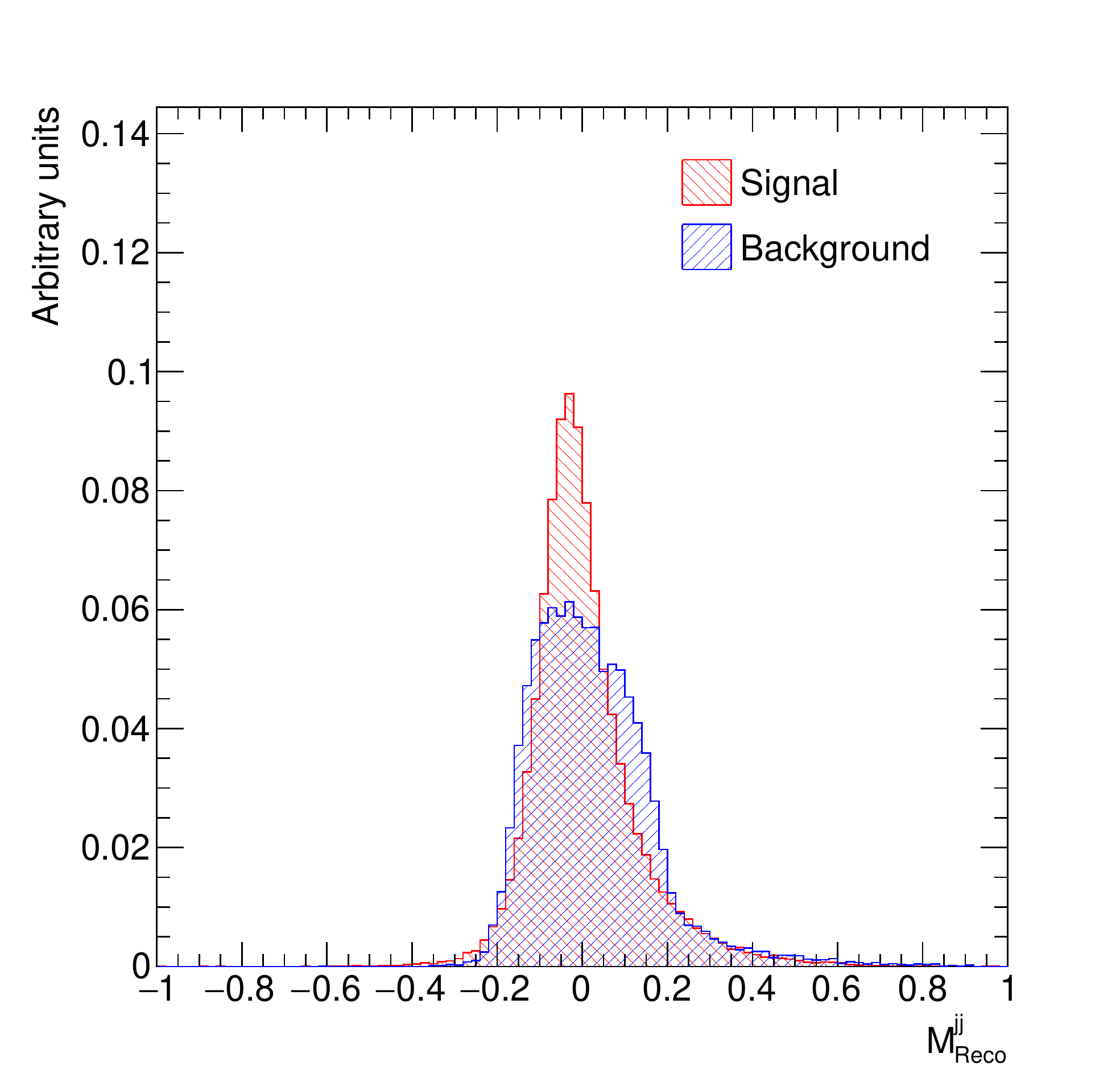}}
  \subfloat[\label{subfig:e_yy}]{\includegraphics[width=0.33\textwidth]{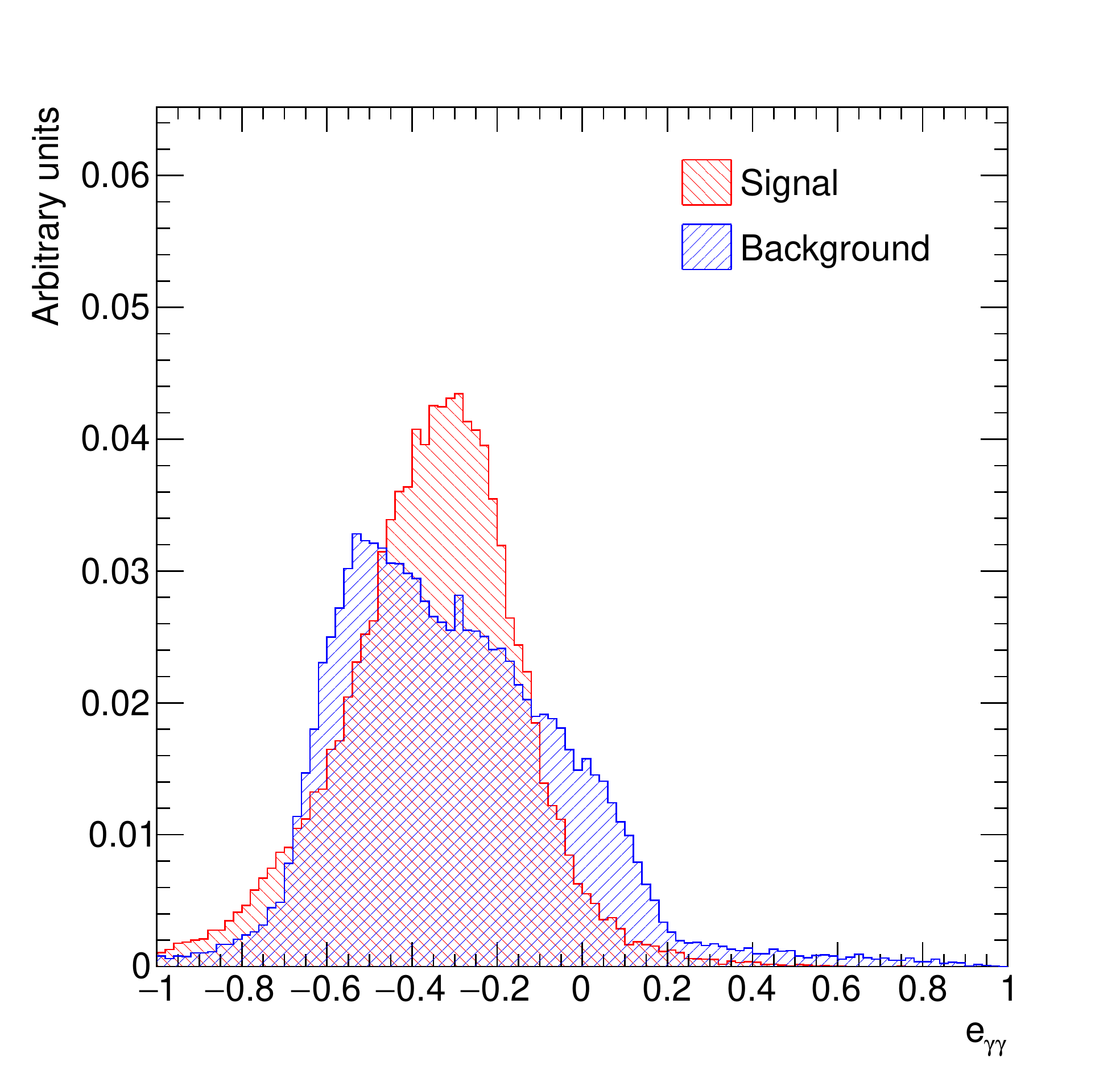}}
  \subfloat[\label{subfig:p_yy}]{\includegraphics[width=0.33\textwidth]{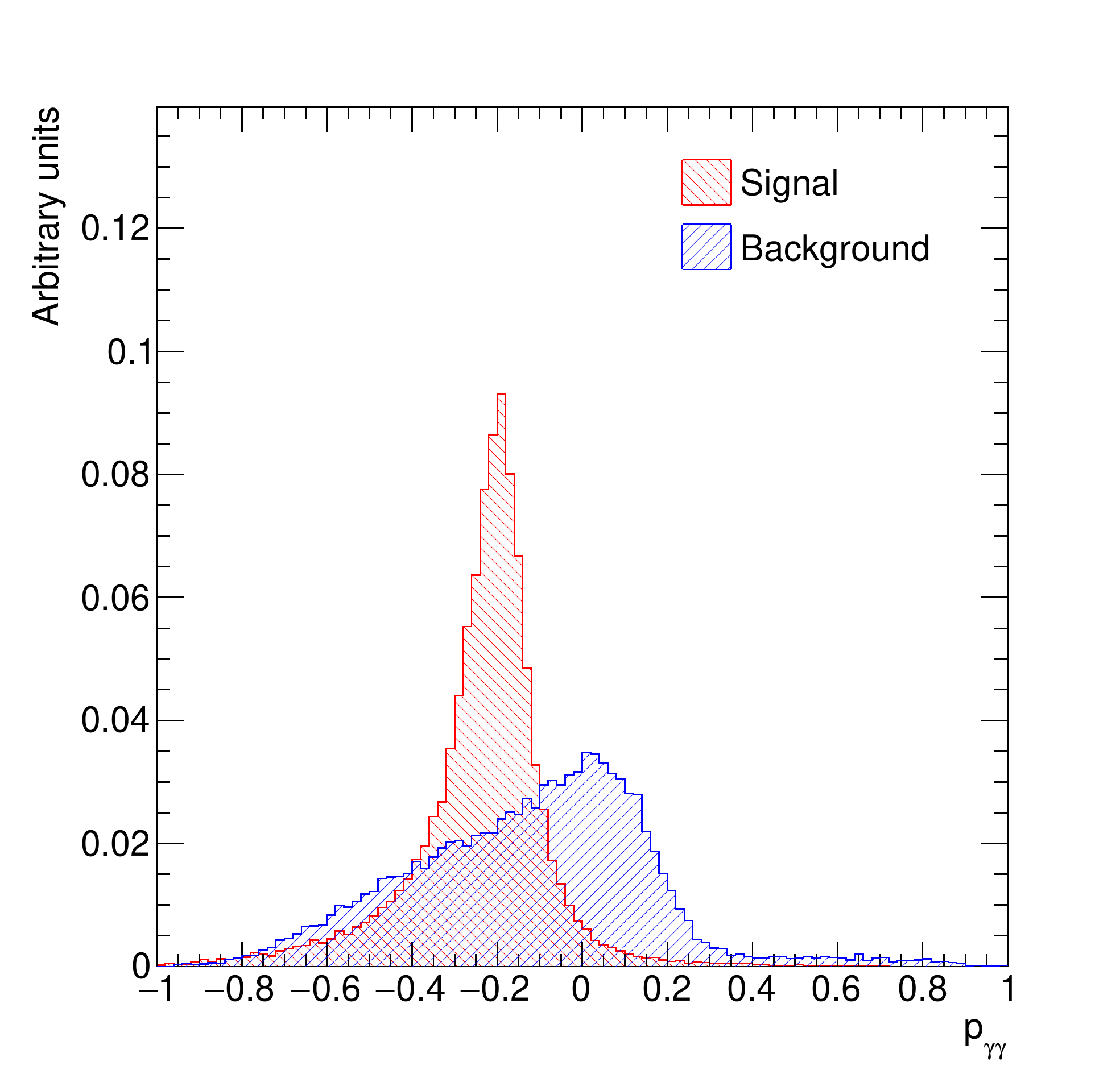}}
  \end{center}
  \caption{Kinematic features for the input variables used in the training and testing. The red histogram represents the $e^+e^-\rightarrow ZH\rightarrow jj\gamma\gamma$ signal, and the blue histogram shows the $e^+e^-\rightarrow (Z/\gamma^*)\gamma\gamma$ SM background process. The $x$-axis for each variable is  to be between -1 and 1.}\label{fig_variables}
\end{figure}

To reduce the size of the background, we first do some pre-selections. The transverse momentum of each photon has to be larger than $20\ \mathrm{GeV}$, and the diphoton invariant mass is required to fall within a broad mass window of $110\ \mathrm{GeV} < m_{\gamma \gamma} < 140\ \mathrm{GeV}$. To separate the signal from the background after pre-selections, we construct classifiers based on (either classical or quantum) machine learning algorithms. The kinematic features utilised by these classifiers include the azimuthal separation between the two photons $\Delta \phi ( \gamma \gamma )$, the minimum angular distance between a photon and a jet $min(\Delta R(\gamma,j))$, energy of the diphoton system $e_{\gamma\gamma}$, the momentum of the diphoton system $P_{\gamma\gamma}$, difference in the momentum between the diphoton system and dijet system $\Delta P_{\gamma\gamma, jj}$, and recoil mass of the dijet system $M^{jj}_{\textrm{recoil}}$.Figure~\ref{fig_variables} shows a comparison between the signal and SM background distributions for these kinematic features. The $x$-axis for each variable is normalised to be between -1 and 1 to enhance the training performance and achieve greater stability. The six variables were chosen from a pool of options based on their superior separate power. The outputs of these classifiers can be used to create event categories and therefore improve the analysis sensitivity. 

\subsection{Quantum algorithm}
The support-vector machine (SVM) algorithm~\cite{svm1,svm} with various kernel estimators has been one of the best-known machine learning algorithms for classification problems, such as identifying a small signal from large backgrounds. A quantum version of the SVM algorithm with a quantum kernel estimator (QSVM-Kernel)~\cite{qsvmv,qsvmm} was proposed to leverage high-dimensional quantum state space for identification accuracy and computational speed. This QSVM-Kernel algorithm is employed and investigated in our study. Conceptually, the training phase of this algorithm can be divided into three sequential steps.

\begin{figure}[!th]
    \begin{center}    
    \subfloat[\label{subfig:qc}]{\includegraphics[width=0.65\textwidth, height=0.2\textheight]{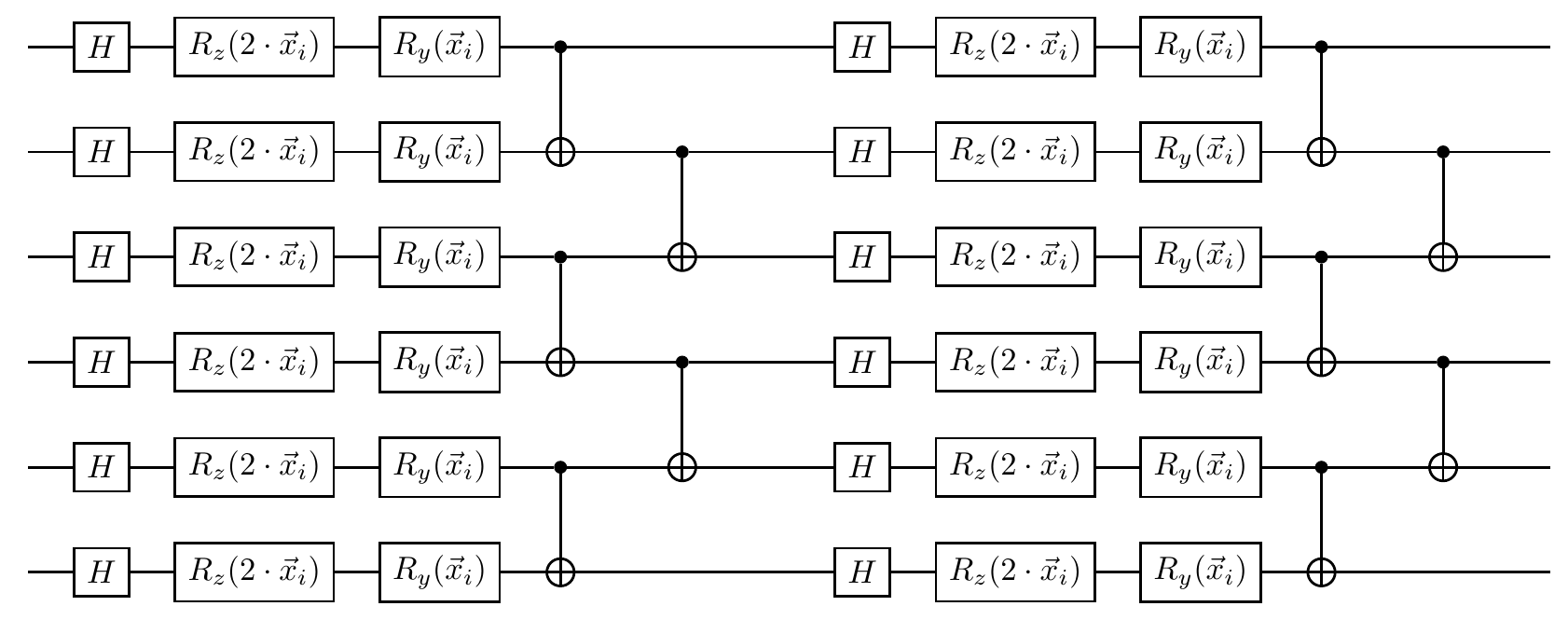}}\\
    \subfloat[\label{subfig:qcval}]{\includegraphics[width=0.65\textwidth, height=0.2\textheight]{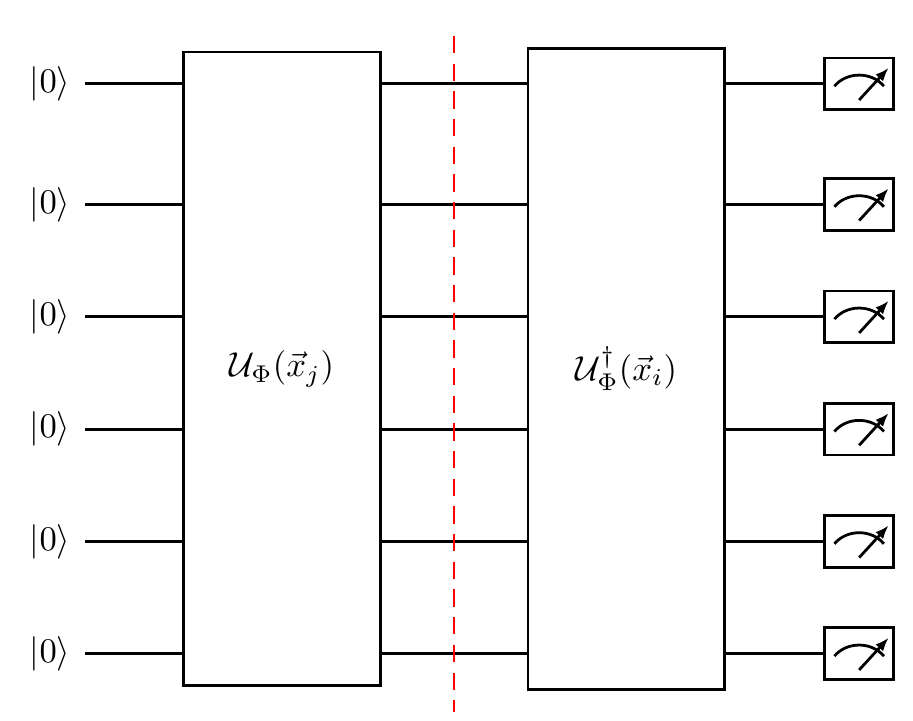}}
    \end{center}
    \caption{\protect\subref{subfig:qc} Quantum circuit of the quantum feature map $\mathcal{U}_{\Phi(\vec{x})}$ in our study. It consists of two identical layers, and each layer is constituted by single-qubit rotation gates ($H$, $R_z$ and $R_y$) and two-qubit CNOT entangling gates.~\protect\subref{subfig:qcval} Quantum circuit for evaluating the kernel entry between two data points $\vec{x_i}$ and $\vec{x_j}$ in our study.}\label{fig_qfmap_qvar}
\end{figure}

\subsubsection{Mapping classical datasets}
\label{Sec:map}
For each classical data point in the training sample, we use $\vec{x}$ to denote the vector of its kinematic features and use $y \in \{-1,1\}$ to denote its event class (-1 for backgrounds and 1 for the signal). The QSVM-Kernel algorithm maps $\vec{x}$ to a quantum state of $N$ qubits, which is a superposition of $2^N$ eigenstates. All the $N$ qubits are initially in the $|0\rangle$ state. A quantum feature map circuit $\mathcal{U}_{\Phi(\vec{x})}$, which represents a unitary transformation, is applied to the $N$ qubits and results in a new quantum state:

\begin{equation}
\left|\Phi(\vec{x})\right\rangle = \mathcal{U}_{\Phi(\vec{x})}\left|0^{\otimes N}\right\rangle 
\end{equation}

The quantum feature map circuit decides how the $2^N$ dimensional quantum state space will be utilised, and the choice of the circuit is essential for the performance of the QSVM-Kernel algorithm. The quantum feature map $\mathcal{U}_{\Phi(\vec{x})}$ in our study, shown in Figure~\ref{fig_qfmap_qvar}~\subref{subfig:qc}, consists of two identical layers. Each layer is constituted by single-qubit rotation gates ($H$, $R_z$ and $R_y$) and two-qubit CNOT entangling gates. Given an input feature vector $\vec{x}$ (where $x_{k}$ denotes the $k^{th}$ element of $\vec{x}$), the $k^{th}$ qubit will be sequentially rotated by an Hadamard ($H$) gate to the $\frac{\left|0\right\rangle + \left|1\right\rangle}{\sqrt{2}}$ state, rotated by a $R_z$ gate for $2\cdot x_{k}$ around the $z$-axis of the Bloch Sphere, and rotated by a $R_y$ gate for $x_{k}$ around the $y$-axis. To avoid long-depth circuits on noisy intermediate-scale quantum computers, CNOT entangling gates (which operate the target qubits according to the state of the control qubits) are arranged in an alternating manner. Firstly, the $k^{th}$ CNOT gate takes the $2\cdot k-1$ qubit as the control qubit and takes the $2\cdot k$ qubit as the target qubit; and secondly, the $k^{th}$ CNOT gate takes the $2\cdot k$ qubit as the control qubit and takes the $2\cdot k+1$ qubit as the target qubit. The best feature map is chosen based on a series of optimisation tests using the training dataset and analysis strategy described in Section~\ref{Simulators results}. The results of the trials, as shown in Table~\ref{tab:optimise}, suggest that rotating twice the feature $\vec{x}_i$ around the $z$ axis and then $\vec{x}_i$ around $y$ axis of the Bloch Sphere has the best AUC value in the 5000 events scan. The entanglement performed via the CNOT gate is also optimised based on the highest AUC value. 

\begin{table}
	\caption{\label{tab:optimise} The table shows optimising the rotation of a single-qubit gate around the Bloch Sphere axes using 5000 events.}
        \centering
	\resizebox{0.8\columnwidth}{!}{%
            \AtBeginEnvironment{tabular}{\tiny}
		\begin{tabular}{|l|c|c|c|c|}
			\hline
			Rotation                 & Depth         & Events                & Best AUC & Variation \\ \hline
			$R_z(2\cdot \vec{x_i})+R_y(\vec{x_i})$    & \multirow{6}{*}{2} & \multirow{6}{*}{5000} & 0.935    & 0.009     \\ \cline{1-1} \cline{4-5} 
			$R_z(\vec{x_i})+R_y(\vec{x_i})$           &                    &                       & 0.933    & 0.015     \\ \cline{1-1} \cline{4-5} 
			$R_y(\vec{x_i})+R_x(\vec{x_i})$           &                    &                       & 0.932    & 0.015     \\ \cline{1-1} \cline{4-5} 
			$R_z(\vec{x_i})+R_z(\vec{x_i})$           &                    &                       & 0.932    & 0.014     \\ \cline{1-1} \cline{4-5} 
			$R_y(\vec{x_i})$                          &                    &                       & 0.928    & 0.008     \\ \cline{1-1} \cline{4-5} 
			$R_z(\vec{x_i})$                          &                    &                       & 0.928    & 0.008     \\ \hline
   		$R_z(2\cdot \vec{x_i})+R_y(\vec{x_i})$    & \multirow{6}{*}{3} & \multirow{6}{*}{5000} & 0.934    & 0.009     \\ \cline{1-1} \cline{4-5} 
			$R_z(\vec{x_i})+R_y(\vec{x_i})$           &                    &                       & 0.932    & 0.014     \\ \cline{1-1} \cline{4-5} 
			$R_y(\vec{x_i})+R_x(\vec{x_i})$           &                    &                       & 0.930    & 0.015     \\ \cline{1-1} \cline{4-5} 
			$R_z(\vec{x_i})+R_z(\vec{x_i})$           &                    &                       & 0.931    & 0.014     \\ \cline{1-1} \cline{4-5} 
			$R_y(\vec{x_i})$                          &                    &                       & 0.927    & 0.008     \\ \cline{1-1} \cline{4-5} 
			$R_z(\vec{x_i})$                          &                    &                       & 0.928    & 0.007     \\ \hline
      	$R_z(2\cdot \vec{x_i})+R_y(\vec{x_i})$    & \multirow{6}{*}{5} & \multirow{6}{*}{5000} & 0.925    & 0.008     \\ \cline{1-1} \cline{4-5} 
			$R_z(\vec{x_i})+R_y(\vec{x_i})$           &                    &                       & 0.923    & 0.015     \\ \cline{1-1} \cline{4-5} 
			$R_y(\vec{x_i})+R_x(\vec{x_i})$           &                    &                       & 0.925    & 0.016     \\ \cline{1-1} \cline{4-5} 
			$R_z(\vec{x_i})+R_z(\vec{x_i})$           &                    &                       & 0.927    & 0.014     \\ \cline{1-1} \cline{4-5} 
			$R_y(\vec{x_i})$                          &                    &                       & 0.925    & 0.006     \\ \cline{1-1} \cline{4-5} 
			$R_z(\vec{x_i})$                          &                    &                       & 0.919    & 0.007     \\ \hline
		\end{tabular}%
	}
\end{table}
\subsubsection{Quantum Kernel estimation}
The QSVM-Kernel algorithm defines the ``kernel entry'' between any two data points $\vec{x_i}$ and $\vec{x_j}$ as the square of the inner product of their quantum states:
\begin{equation}
    k(\vec{x_i}, \vec{x_j}) = \left|\left\langle\Phi(\vec{x_i})\right|\left.\Phi(\vec{x_j})\right\rangle\right|^2. 
\end{equation}
The mathematical implication of the kernel entry is the distance between the two data points in the high-dimensional quantum state space. It can be shown that:
\begin{equation}
    k(\vec{x_i}, \vec{x_j}) 
    = \left|\left\langle 0^{\otimes N}\right|\mathcal{U}^{\dag}_{\Phi(\vec{x_i})}\mathcal{U}_{\Phi(\vec{x_j})}\left|0^{\otimes N}\right\rangle\right|^2.
\end{equation}
Therefore the kernel entry can be calculated using $N$ qubits on a quantum computer by preparing the 
$\mathcal{U}^{\dag}_{\Phi(\vec{x_i})}\mathcal{U}_{\Phi(\vec{x_j})}\left|0^{\otimes N}\right\rangle$ state and measuring it on a standard basis with a sufficient number of measurement shots. The frequency of obtaining the $|0^{\otimes N}\rangle$ state in the measurement outputs is taken as the value of the kernel entry. Except for the $\mathcal{U}_{\Phi(\vec{x})}$ design, the quantum circuit for calculating the kernel entries is the same as Ref~\cite{qsvmv} and given in Figure~\ref{fig_qfmap_qvar}~\subref{subfig:qcval}.

\subsubsection{Finding separating hyperplane}
Using the kernel entries, the QSVM-Kernel algorithm looks for a hyperplane that separates the signal from the background in the quantum state space:

\begin{equation}
     \sum_{i=1}^{t} \alpha_i y_i k(\vec{x}_i, \vec{x}) + b = 0   
\end{equation}
where $t$ is the size of the training dataset, $i$ is the index of the training data points,
and $(\alpha_i, b)$ are parameters to be optimised. The optimisation of the separating hyperplane takes the same procedure as for the classical SVM and is done in a classical computer. 

In the testing phase of the QSVM-Kernel algorithm, given any new data point $\vec{x}'$,
the kernel entry between $\vec{x}'$ and each training data point is calculated on a quantum computer. Then, on a classical computer and as for the classical SVM, the data point $\vec{x}'$ is classified as ``signal'' or ``background'' based on the separating hyperplane, i.e. the sign of $(\sum_{i=1}^{t} \alpha_i y_i k(\vec{x}_i, \vec{x}') + b)$. In addition, the probability for the data point $\vec{x}'$ to be in the signal class can be evaluated and used as a continuous discriminant.
\section{Results}
\subsection{Results from Quantum Computer Simulators}
\label{Simulators results}
\begin{figure}[htb]
  \begin{center}
    \includegraphics[width=0.7\textwidth]{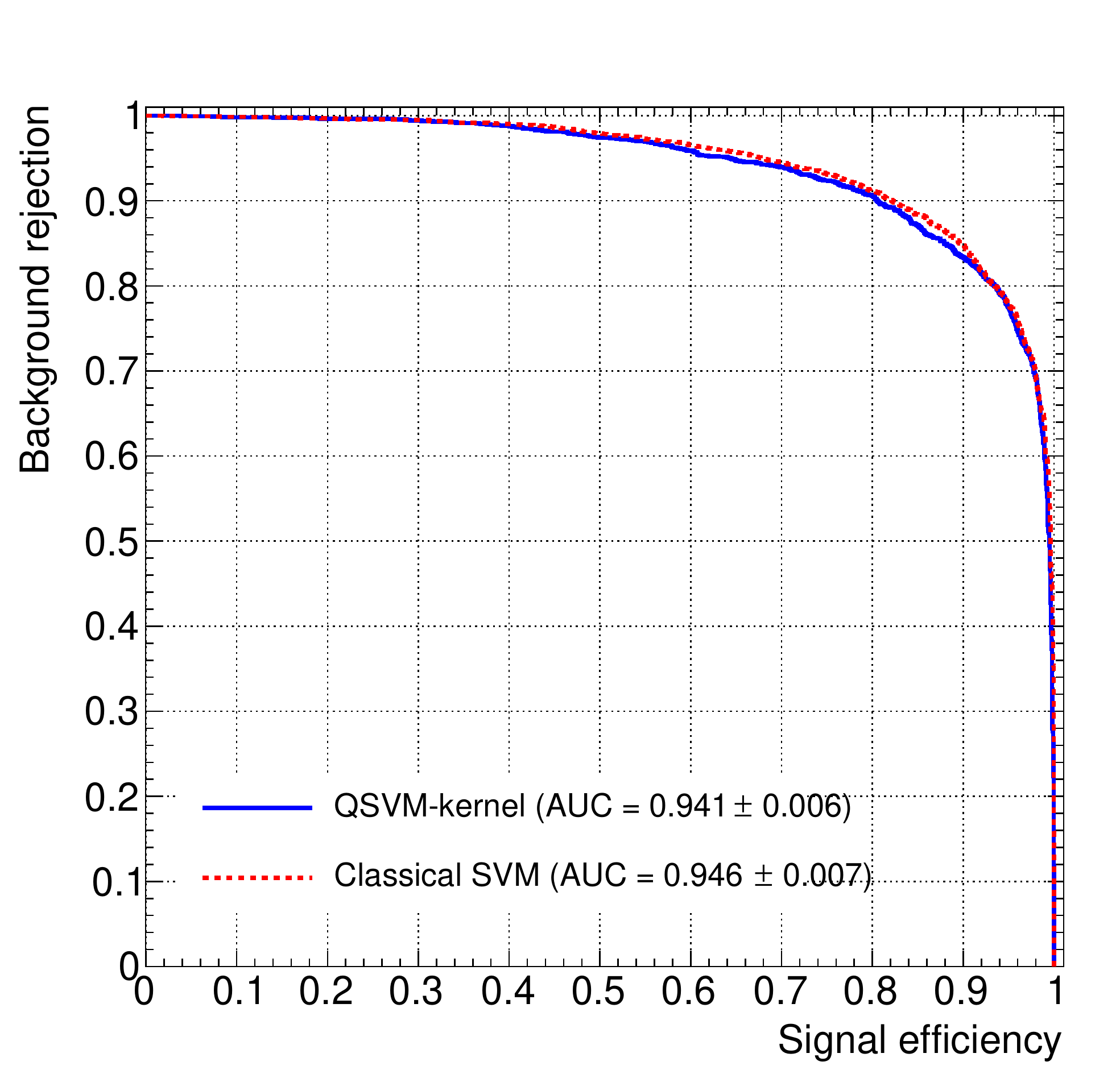}
   \end{center}
     \caption{ROC curves of machine learning classifiers using the $e^{+}e^{-} \rightarrow ZH$ analysis dataset with 12000 events. The plot overlays the results of the QSVM-Kernel algorithm (blue) and the classical SVM algorithm (red).}\label{fig2}
\end{figure}

\begin{figure*}
	\begin{center}
	\subfloat[\label{subfig:auc_events}]{\includegraphics[width=0.48\textwidth]{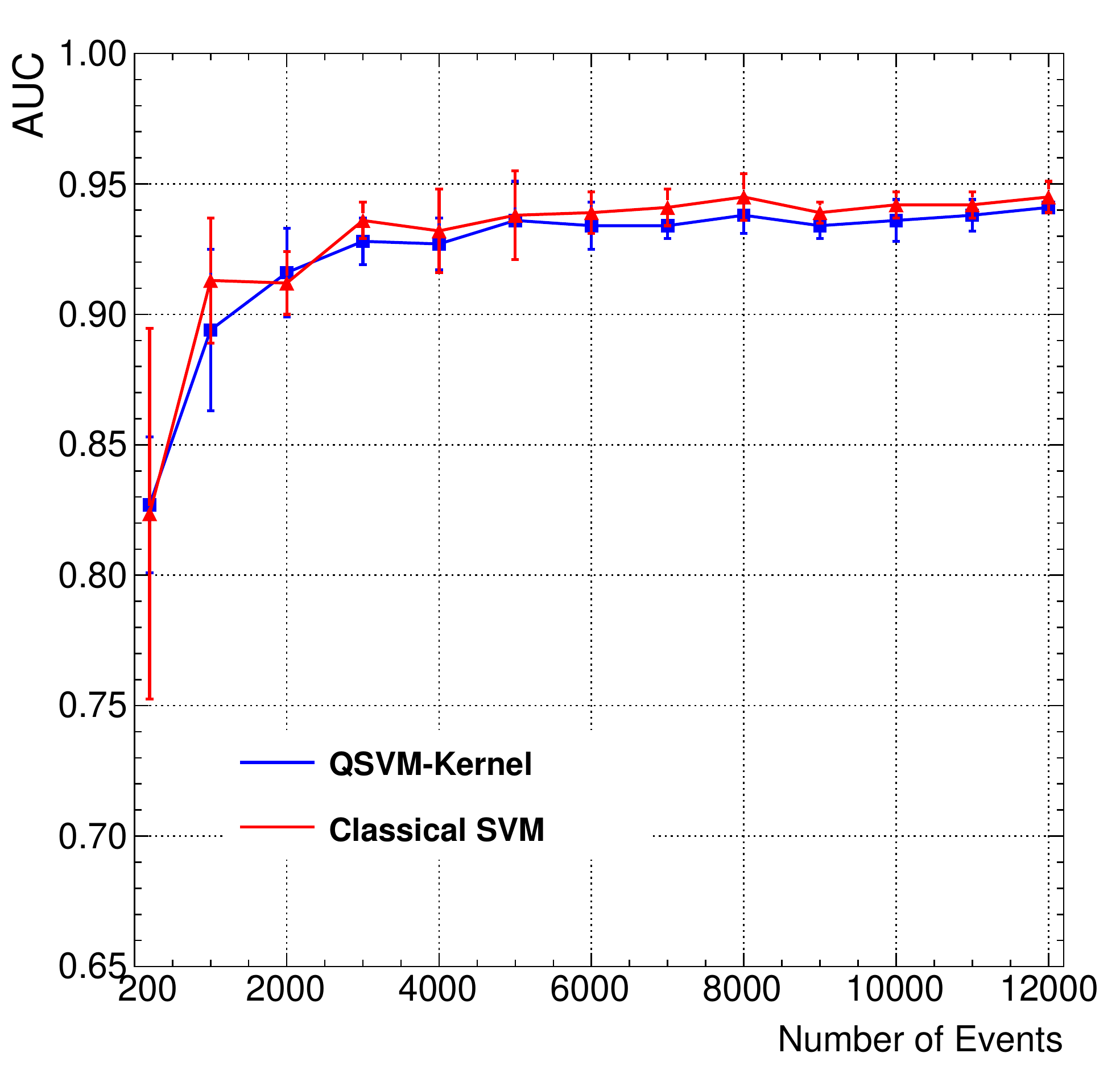}}
        \quad
	\subfloat[\label{subfig:del_auc_events}]{\includegraphics[width=0.48\textwidth]{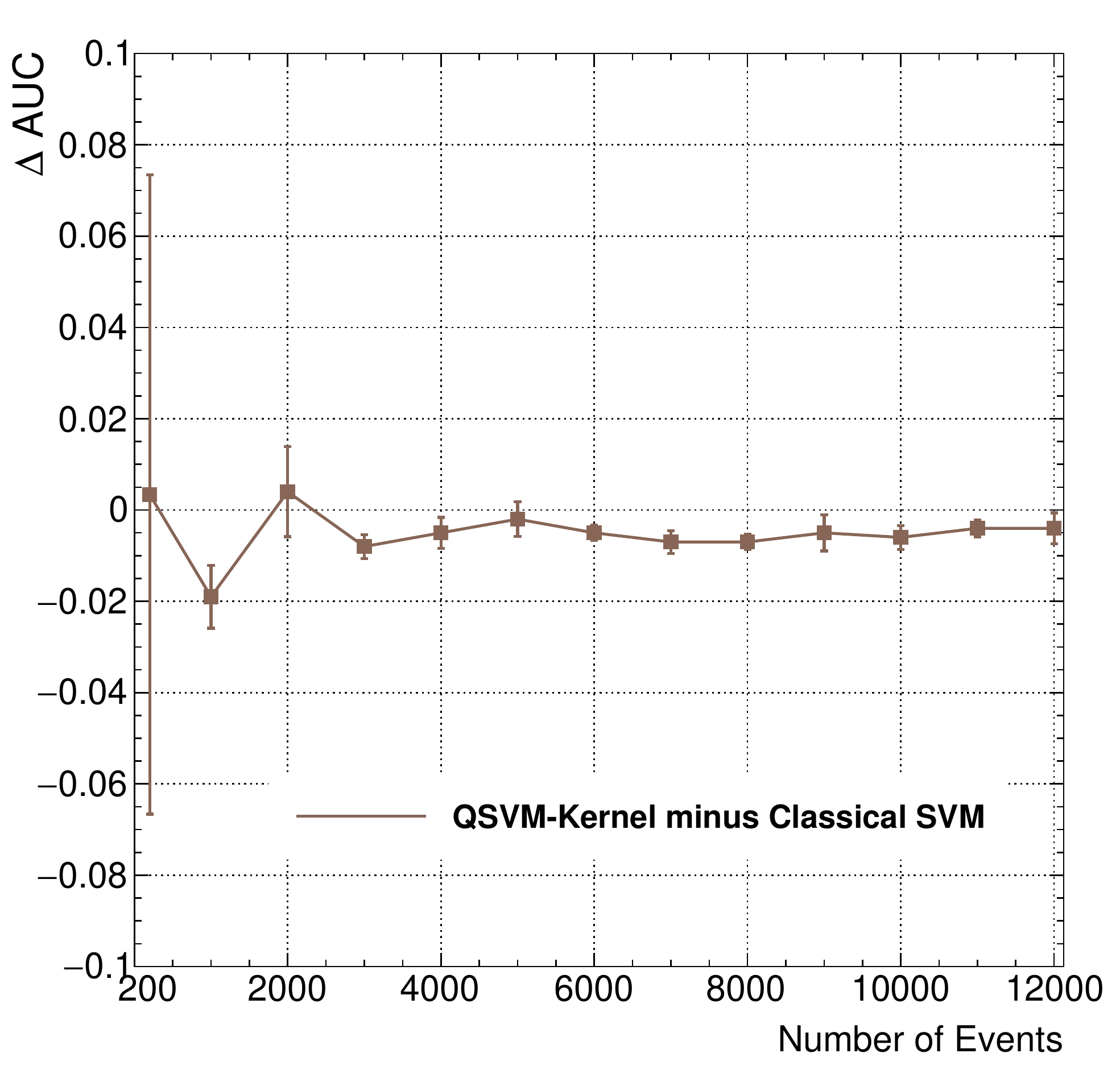}}
        \end{center}
	\caption{The AUCs of the QSVM-Kernel and classical SVM classifiers for the $e^{+}e^{-} \rightarrow ZH$ analysis with different dataset sizes of 200 and from 1000 to 12000 events. Figure~\protect\subref{subfig:auc_events} overlays the AUC results of the QSVM-Kernel and the classical SVM as a function of the dataset size. Figure~\protect\subref{subfig:del_auc_events} shows the difference in AUC between the QSVM-Kernel algorithm and the classical SVM algorithm, again as a function of the dataset size. The quoted AUCs are the mean for the AUCs of several shuffles for the dataset, and the quoted errors are the standard deviations for the AUCs of the shuffles.}\label{fig3}
\end{figure*}

We employ the QSVM-Kernel algorithm using 6-qubits on the \textit{StatevectorSimulator} from the IBM Quantum framework~\cite{qiskit} and the \textit{Tensor Network Simulator} from the Origin \textsc{QPanda} framework~\cite{qpanda}. Running on classical computers, these tools simulate quantum computers by computing the wave function of the qubits as the quantum gates are executed
and output the ``true'' probabilities for each eigenstate when measurements are performed. In our usage of quantum computer simulators, hardware noises are not considered. These choices lead to reasonable computational speed that allows us to explore large datasets. For a $e^{+}e^{-} \rightarrow ZH$ analysis dataset of size $M$, we prepare $\frac{1}{2}M$ simulated signal events and $\frac{1}{2}M$ simulated background events. 
The dataset is divided into a training sample and a test sample. The training sample comprises $\frac{1}{2}M$ simulated signal and background events, while the test sample contains the remaining half of the events, with an equal distribution of signal and background events (50:50).
To evaluate the statistical fluctuation level, we perform this splitting several times, train and test a QSVM-Kernel classifier for each shuffle, and report the average and variation of the results. The SVM regularisation hyperparameter for QSVM-Kernel is optimised using a cross-validation procedure~\cite{cv1,cv2}. With the same datasets, a classical SVM~\cite{svm1,svm} classifier with the RBF kernel is trained using the scikit-learn package~\cite{sklearn}, which serves as the classical counterpart for the QSVM-Kernel classifier. Using the same cross-validation procedure as for QSVM-Kernel, we extensively optimise the SVM regularisation hyperparameter and the RBF kernel's $\gamma$ parameter.

We first build QSVM-Kernel and classical SVM classifiers for a $e^{+}e^{-} \rightarrow ZH$ analysis dataset of 12000 events. To study the performance of the machine learning-based models, we plot Receiver Operating Characteristic (ROC) curves, which show background rejection (in $y$-axis) as a function of signal efficiency (in $x$-axis) using the continuous discriminants of the classifiers.
The ROC curves for the QSVM-Kernel algorithm (blue) and the classical SVM algorithm (red) are overlaid in Figure~\ref{fig2}. The IBM \textit{StatevectorSimulator} and the Origin \textit{Tensor Network Simulator} produce identical ROC curves and are therefore shown together. Additionally, we calculate the areas under the ROC curves (AUCs) for the classifiers, which is a quantitative metric for evaluating the performances of machine learning applications. The AUC of the QSVM-Kernel classifier is found to be 0.941$\pm$0.006, compared to 0.946$\pm$0.007 for the classical SVM classifier. The comparison with the ROC curves and AUCs indicates that the quantum SVM algorithm provides similar separation power with its classical counterpart in the $e^{+}e^{-} \rightarrow ZH$ analysis with the same dataset having 12000 events.

Furthermore, we construct QSVM-Kernel and classical SVM classifiers for the $e^{+}e^{-} \rightarrow ZH$ analysis with different dataset sizes from 1000 to 12000 events. Figure~\ref{fig3}~\subref{subfig:auc_events} overlays the AUC results of the QSVM-Kernel and the classical SVM as a function of the dataset size. Figure~\ref{fig3}~\subref{subfig:del_auc_events} further shows the difference in AUC between the QSVM-Kernel algorithm and the classical SVM algorithm, again as a function of the dataset size. The quoted AUCs are the mean for the AUCs of several shuffles for a dataset and the quoted errors are the standard deviation for the AUCs of the shuffles. As shown in the figure, concerning separating signal from background for the $e^{+}e^{-} \rightarrow ZH$ analysis, both algorithms improve the performances with larger dataset size, and for up to 12000 events, the QSVM-Kernel algorithm performs similarly to the classical SVM algorithm.

\begin{figure}[!t]
  \begin{center}
   \subfloat[\label{subfig:ibm_hard}]{\includegraphics[width=0.7\textwidth]{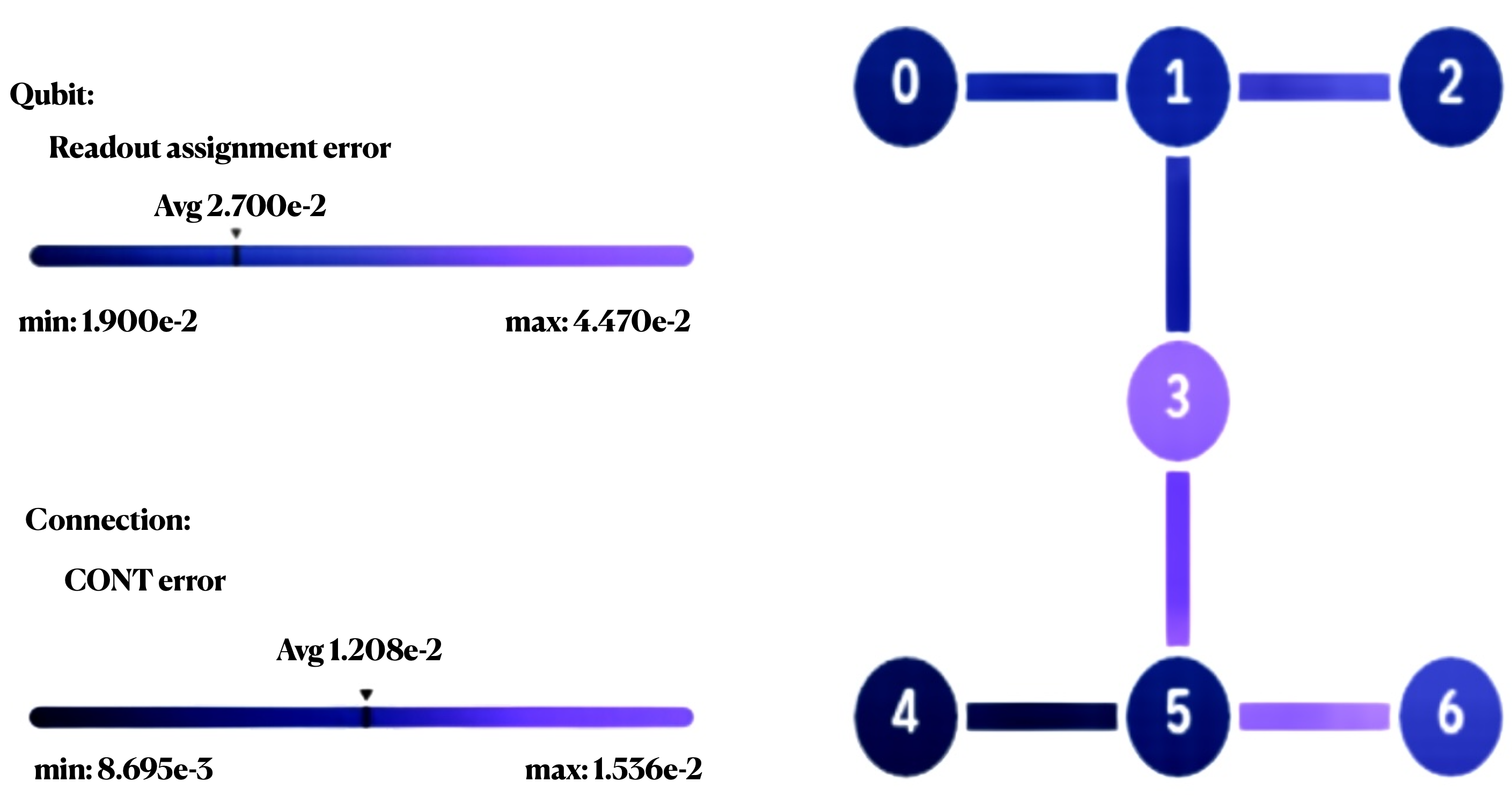}}\\
   \subfloat[\label{subfig:wuyuan_hard}]{\includegraphics[width=0.7\textwidth]{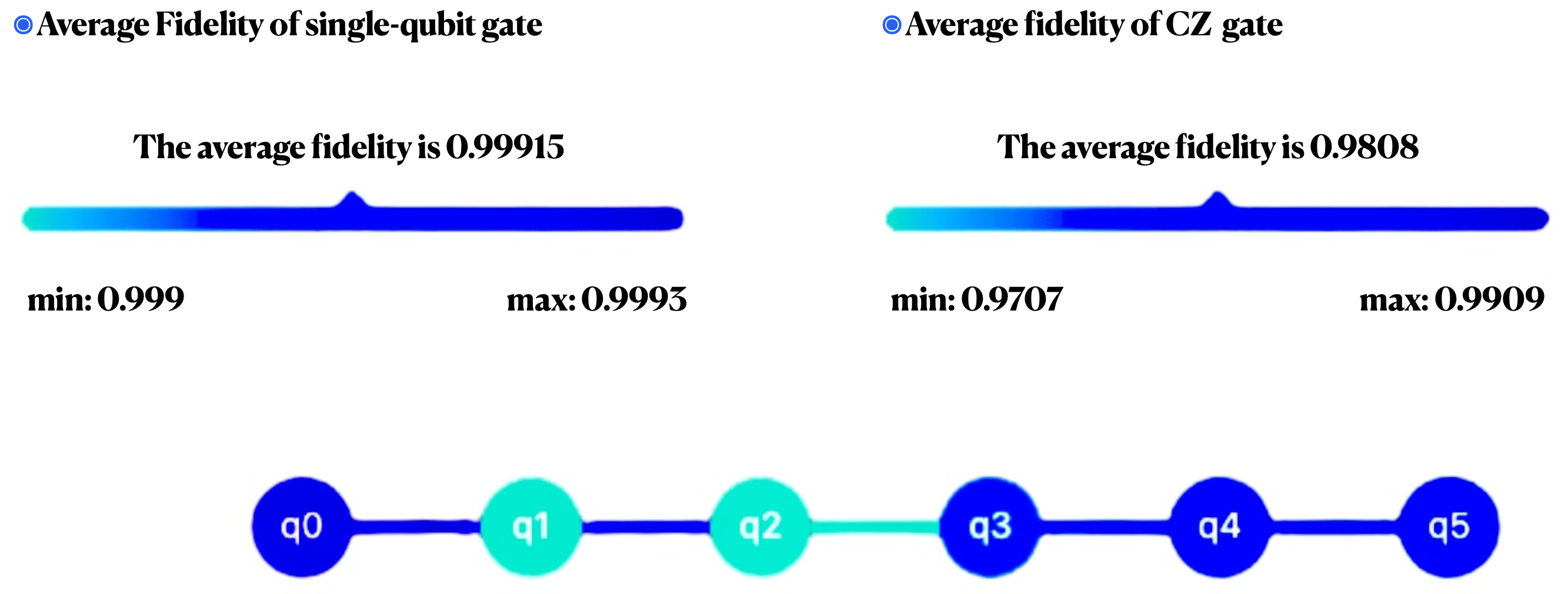}}
   \end{center}
   \caption{The topology structure of the~\protect\subref{subfig:ibm_hard} 7-qubits in the IBM Nairobi and~\protect\subref{subfig:wuyuan_hard} 6-qubits in the Origin Wuyuan quantum chip systems. 
     For~\protect\subref{subfig:ibm_hard}, our study uses qubit 0, 1, 2, 3, 4, and 5.}\label{fig_wuyuan_qubits}
\end{figure}

\begin{figure}[htb]
	\begin{center}
	\includegraphics[width=0.7\textwidth]{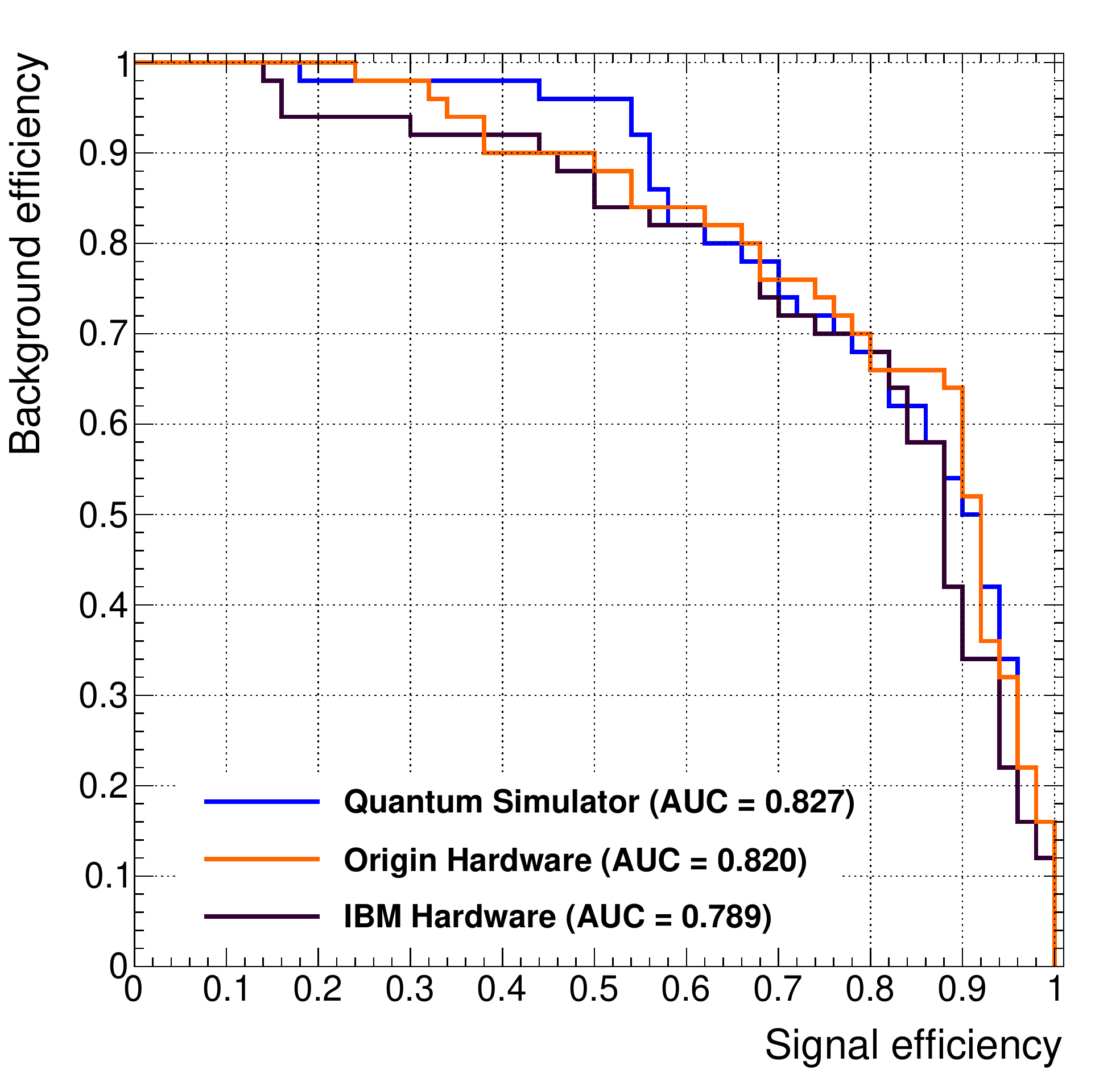}
        \end{center}
	\caption{The ROC curves of the QSVM-Kernel classifiers from the IBM Nairobi quantum computer hardware, the Origin Wuyuan quantum computer hardware and the quantum computer simulators with the same $e^{+}e^{-} \rightarrow ZH$ analysis dataset.}\label{fig4}
\end{figure}

\subsection{Results from Quantum Computer Hardware}
In addition to the studies with noiseless quantum computer simulators, we further investigate the QSVM-Kernel algorithm with noisy quantum computer hardware, the 7-qubit Nairobi quantum computer from the IBM company and the 6-qubit Wuyuan quantum computer from the Origin Quantum company. The core of both quantum computers is a superconducting quantum chip system based on the Josephson Junction Structure. We employed 6 qubits on both the IBM Nairobi and Origin Wuyuan quantum computers. The topology structures of the qubits in the quantum chip systems are shown in Figure~\ref{fig_wuyuan_qubits}. 
These computers have an identical quantum circuit configuration as used in the simulation.
Each quantum circuit was executed and measured 10000 times to allow sufficient statistical precision in evaluating kernel entries. Due to the currently long execution time of the quantum computer hardware, here we analyse a smaller $e^{+}e^{-} \rightarrow ZH$ analysis dataset consisting of 100 training events and 100 test events.

We plot the ROC curves of the QSVM-Kernel classifiers from the IBM Nairobi and Origin Wuyuan quantum computer hardware, as shown in Figure~\ref{fig4}.
The ROC curve from the \textit{StatevectorSimulator} and \textit{Tensor Network Simulator} with the same $e^{+}e^{-} \rightarrow ZH$ analysis dataset overlaid for comparison. The AUC reaches 0.820 for the Origin Wuyuan quantum computer hardware and 0.789 for the IBM Nairobi quantum computer hardware, compared to 0.827 for the quantum computer simulators. We find that the separation power provided by the current quantum computer hardware is approaching that by the noiseless quantum computer simulators. The remaining differences between IBM Nairobi's and Origin Wuyuan's results are related to the quantum hardware noises and statistical fluctuations while executing the hardware tasks. Also, the difference between IBM Nairobi and simulation results is due to the hardware noise and statistical uncertainty. The IBM Quantum framework automatically generates a simplified noise model for an actual device. In our case, the noise model is forged with \textit{FakeNairobi}, which uses calibration information reported in the backend properties of IBM Nairobi. The noise model considers the gate error probability, readout error probability, and $T_1$ and $T_2$ relaxation time constant of each qubit~\cite{qiskit}. The noise model is then incorporated into the IBM QASM simulator. The estimated noise on the ROC curve is 0.017. Notice that this value is calculated based only on IBM and not on Origin quantum hardware. The standard deviation of results generated using different seeds is 0.022, which is taken as statistical fluctuations. The IBM hardware and simulator results are consistent within the margin of the uncertainties.  

\section{Discussion}
In this study, we have employed the QSVM-Kernel algorithm to study the $ZH$ process (Higgs boson production in association with a $Z$ boson) at the CEPC, a proposed Higgs factory for exploring the deeper structure of particle physics. QSVM-Kernel, a quantum machine learning algorithm, can leverage high-dimensional quantum state space for identifying a signal from backgrounds. Using 6-qubits on quantum computer simulators, we optimised the QSVM-Kernel algorithm's quantum circuits for our particle physics data analysis. We obtained a classification performance similar to the classical SVM algorithm with different dataset sizes (of 200 and from 1000 to 12000 events). Furthermore, we have validated the QSVM-Kernel algorithm using 6 qubits on superconducting quantum computer hardware from both IBM and Origin Quantum. This validation characterises the classification performance for a small dataset (of 100 events), which approaches noiseless quantum computer simulators. Our study is one of the examples that apply state-of-the-art quantum computing technologies to particle physics, a branch of fundamental science that relies on big experimental data.

\begin{itemize}
    \item For the first time, we employed quantum machine learning in physics studies for future Higgs factories. Higgs factories, including FCC-ee, CEPC, etc., will produce a large number of Higgs boson events to study the structure and origin of particle physics in-depth. When these Higgs factories start to produce collisions, much more powerful quantum computer hardware will probably exist compared to what we have today.
    We expect improved quantum machine learning algorithms and quantum computer hardware could facilitate physics discovery in the exciting data planned to be produced by the CEPC. 
    Therefore, preparing quantum algorithms for the Higgs factories is valuable. 

    \item We are using the QSVM-Kernel algorithm first published in 2018~\cite{qsvmv,qsvmm}. The core part of this algorithm is the quantum feature map, which encodes the classical data into a quantum state. We worked extensively to optimise the quantum feature map for our physics study. Finally, we obtained a feature map different from other publications (e.g. ref~\cite{qsvmtth}) that works best in our physics case.

    \item For the first time, we compared results based on  IBM and Origin Quantum quantum computer hardware by utilising quantum machine learning in particle physics. This is a very valuable comparison: our result demonstrated both quantum computer hardware could utilize quantum state space built with realistic particle physics datasets. We had to work intensively to commission our quantum study for both IBM and Origin Quantum resources. 
\end{itemize}

Looking toward the future, given the incredible momentum in the field of quantum technology, we expect to see significantly improved quantum machine learning algorithms and quantum computer hardware with more qubits and better computational speed. 
With these, particle physicists could better handle high dimensions and large-size physics datasets from future particle colliders, providing possibilities for discovering new fundamental particles and interactions. The QSVM-Kernel algorithm can process a larger number of features without increasing computational time if the number of features is smaller than the number of qubits. This indicates that QSVM can better handle complicated final states at future particle colliders. In addition, the QSVM-Kernel algorithm scales quadratically with the size of the dataset. The algorithm complexity is $O(n^2)$ because we need to calculate the entire $N\times N$ kernel matrix as the classical SVM. This is not ideal but can be mitigated by the better computational speed of future quantum computer hardware. We plan to improve the current QSVM-Kernel algorithm to overcome this challenge. 

\section*{Acknowledgements}
This research is funded by the NSFC Basic Science Centre Program for “Joint Research on High Energy Frontier Particle Physics” (Grant No. 12188102), 
the Science and Technology Innovation Project of the Institute of High Energy Physics, Chinese Academy of Sciences (Grand No. E25453U210),
the State Key Laboratory of Nuclear Physics and Technology, Peking University (Grant No. NPT2022ZZ05),
and the Fundamental Research Funds for the Central Universities, Peking University.
The authors acknowledge the computing and data resources provided by the computing centre of the Institute of High Energy Physics, National High Energy Physics Data Centre, Chinese Academy of Science.
The authors thank IBM and Origin Quantum for providing quantum computer simulators and hardware.
The authors thank Dr. Teng Li (Shangdong University), Fangyi Guo (IHEP), and Congqiao Li (Peking University) for their helpful discussion.

\section*{Conflict of interest statement}
On behalf of all authors, the corresponding author states that there is no conflict of interest.

\end{document}